\RequirePackage{fix-cm}

\documentclass[a4paper,12pt]{article}

\DeclareMathSymbol{\Psi}{\mathalpha}{operators}{9}
\usepackage{graphicx}
\usepackage{amsmath,amsfonts,mathtools}
\usepackage[utf8]{inputenc}
\usepackage[cal=boondoxo]{mathalfa}
\usepackage{bm}
\usepackage{upgreek}
\usepackage{mathptmx}      
\usepackage[symbol]{footmisc}

\DeclareSymbolFont{newfont}{OML}{cmm}{m}{it}
\DeclareMathSymbol{\epsilon}{3}{newfont}{15}

\newcommand{\trans}{^{\mathrm{T}}}
\newcommand{\dM}{\partial M}

\newcommand{\vect}[1]{\mathbf{#1}}
\newcommand{\bb}[1]{\mathbf{#1}}
\newcommand{\bs}[1]{\bm{#1}}

\newcommand{\inprod}[2]{\langle#1\,,\,#2\rangle}

\newcommand{\diff}{\mathrm{d}}

\newcommand{\intUtilde}[1]{\int_{\tilde{U}}#1\,\mathrm{d}\tilde{V}}
\newcommand{\intdUtilde}[1]{\int_{\partial \tilde{U}}#1\,\mathrm{d}\tilde{S}}
\newcommand{\intU}[1]{\int_U#1\,\mathrm{d}V}
\newcommand{\intdU}[1]{\int_{\partial U}#1\dS}

\newcommand{\ofx}{\!\left(\mathbf{x}\right)}
\newcommand{\ofxt}{\!\left(\mathbf{x},t\right)}

\newcommand{\xix}{\bm{\xi}(\vect{x})}

\newcommand{\ofxixt}{[\bm{\xi}(\mathbf{x}),t]}
\newcommand{\ofxix}{[\bm{\xi}(\mathbf{x})]}

\renewcommand{\div}{\mathrm{div}}
\newcommand{\Div}{\mathrm{Div}\,}

\newcommand{\pderiv}[2]{\frac{\partial#1}{\partial#2}}

\newcommand{\Fu}{\mathbf{F}_{u}}

\newcommand{\dS}{\,\mathrm{d}S}
\newcommand{\vx}{\vect{x}}
\newcommand{\Fxi}{\vect{F}_{\xi}}
\newcommand{\FxiT}{\vect{F}_{\xi}^{\mathrm{T}}}
\newcommand{\Fxiinv}{\vect{F}_{\xi}^{-1}}
\newcommand{\FxiinvT}{\vect{F}_{\xi}^{-\mathrm{T}}}

\newcommand{\vphi}{\bm{\varphi}}
\newcommand{\vq}{\vect{q}}
\newcommand{\Th}{\vartheta}
\newcommand{\vT}{\vect{T}}
\newcommand{\vF}{\vect{F}}
\newcommand{\args}{(\vect{F},\Th, \nabla \Th, \vx)}

\newcommand{\vv}{\vect{v}}
\newcommand{\vn}{\vect{n}}
\newcommand{\vb}{\vect{b}}
\newcommand{\vu}{\vect{u}}
\newcommand{\elast}{\bm{\upLambda}}
\newcommand{\vM}{\vect{M}}
\newcommand{\vk}{\vect{k}}
\newcommand{\comp}[1]{#1\circ\bm{\xi}}
\newcommand{\bcomp}[1]{(#1\circ\bm{\xi})}
\newcommand{\zero}{^{(0)}}
\newcommand{\one}{^{(1)}}
\newcommand{\two}{^{(2)}}

\begin{document}

\title{Thermoelastic cloaking and the breaking of invisibility
}


\author{F. Syvret$^1$\footnote{Corresponding author, \texttt{fs406@cam.ac.uk}}   \,   \&
  D. Al-Attar$^1$ \\
  \footnotesize 1 \textit{Bullard Laboratories, Department of Earth Sciences, University of Cambridge, Cambridge, UK} \\
}



\date{Month date, 2019}

\maketitle

\begin{abstract}
  The invariance under co-ordinate transformations of a set of partial differential equations can lead to the possibility of invisibility cloaking, whereby an anomaly in the interior of a body is shielded from an external observer. The form invariance of the equations of thermoelasticity under a class of transformations between reference configurations is demonstrated, with implications for thermal and thermoelastic invisibility cloaking discussed. Specifically, the theory of thermoelasticity is found to not allow for invisibility cloaking, similarly to elastodynamics. Thermal invisibility cloaking is therefore broken by coupling to mechanics. In the case of weakly coupled linear thermoelasticity, measurements of temperature by external observers are shown to be perturbed by the weak thermomechanical coupling, but only to second order in the degree of coupling. This implies the phenomenon of approximate invisibility, meaning that a thermal cloaking device with weak coupling to mechanics could maintain near-perfect cloaking.
\end{abstract}


%
%

%
%
%
%
\section{Introduction}\label{sec:introduction}
The partial differential equations governing several classical field theories are known to be form
invariant under active co-ordinate transformations provided that the fields and parameters of
the theory are simultaneously transformed in an appropriate manner.
Such form invariance is closely related to the prospect of invisibility   cloaking, i.e. concealing
exactly the interior properties of a body from an external observer of the fields. This topic has been most extensively studied in the case of electromagnetism \cite{Pendry2006,Leonhardt2006,Greenleaf2008,Greenleaf2009}. The form invariance property of Maxwell's equations allows, in principle, for the creation of cloaking devices to perfectly conceal an object. These can be practically realised through the use of artificial metamaterials to approximate the  required electromagnetic properties (e.g. \cite{Schurig2006,Chen2008}).

Form invariance and the possibility of  cloaking in the context of  elastodynamics has received
comparatively less attention within theoretical work.  One approach to cloaking for linear elasticity was presented in \cite{Milton2006} and developed in \cite{Norris2011}, though this  required the introduction of non-standard constitutive  relations that violate the principle of material frame indifference.  Within
the context of finite elasticity, form invariance of the governing equations
has been discussed in \cite{Mazzucato2006}, \cite{Wang2019}, \cite{Al-Attar2016} and \cite{Yavari2018}, 
with the results of the latter two papers making clear that the
form invariance does not, in this case, allow for invisibility cloaking.
In brief, this is because the transformations under which the equations are form invariant simply relate two different referential descriptions of the same physical body.

The aim of this work is to investigate the form invariance properties of finite thermoelasticity. The heat diffusion equation is known to be form invariant, and thermal invisibility cloaking is, in principle, feasible \cite{Guenneau2012}. Any real-world thermal cloaking devices will be subject to (perhaps small) mechanical deformations, and it is unknown how this might impact the performance of thermal cloaks.
In  Sect. \ref{sec:heat} we recall briefly the origin of thermal invisibility for the heat diffusion equation. The form invariance of finite thermoelasticity is demonstrated in Sect. \ref{sec:thermoelasticity},
with the results generalising those obtained in \cite{Al-Attar2016} in
the context of finite elasticity. In Sect. \ref{sec:linthermo}  the linearisation of the theory for
small thermal and mechanical disturbances
is considered, with this being useful for understanding the effects of mechanical deformation
upon thermal invisibility.

\section{Form invariance and the heat equation}\label{sec:heat}

To illustrate the  meaning of  form invariance and its relationship to invisibility, it will be
useful to start with the  familiar example of the heat diffusion equation.
Working for simplicity with a  steady-state problem, the  anisotropic
heat equation on a suitably regular domain $M\subseteq \mathbb{R}^{n}$  is given by
\begin{align}\label{eq:heat_differential}
  \div\big(\bb{k}\cdot\nabla \Th\big) &= 0,
\end{align}
where  $\Th$ is the temperature field, $\bb{k}$ is the  conductivity tensor which
takes values in the set of  symmetric and positive-definite linear operators on $\mathbb{R}^{n}$,
and the action of a linear operator on a vector is indicated by a dot.
All fields are assumed to possess appropriate regularity properties, while we
write $\div$ and $\nabla$ for the  divergence and gradient
operators on $\mathbb{R}^{n}$, respectively.
On the boundary $\partial M$ of the domain we apply insulating boundary condition
\begin{align}\label{eq:heat_differential_bnd}
  \inprod{\bb{k}\cdot\nabla \Th}{\bb{n}} &= 0, 
\end{align}
where $\bb{n}$ is the unit outward normal vector, and we write $\inprod{\cdot}{\cdot}$
for the Euclidean inner product on $\mathbb{R}^{n}$.

A simple way to arrive at the form invariance properties for this problem starts by first transforming
the governing equations into an equivalent integral statement of the first law of thermodynamics.
In a steady state, the internal energy contained in an arbitrary sub-domain $U\subseteq M$, with boundary $\partial U$, must be constant, and so the net heat flux out of the sub-domain vanishes:
\begin{equation}\label{eq:heat_integral}
  \intdU{\inprod{\bb{k}\cdot\nabla \Th}{\bb{n}}} = 0.
\end{equation}
Since the sub-domain $U$ is arbitrary, an application of the divergence theorem shows readily
that this condition is equivalent to the heat diffusion equation \eqref{eq:heat_differential} along with the insulating boundary
condition \eqref{eq:heat_differential_bnd}.

Consider a second domain $\tilde{M}\subseteq \mathbb{R}^{n}$ and an
orientation-preserving diffeomorphism $\bm{\xi} : \tilde{M} \rightarrow M$, this being a mapping  which
is smooth, smoothly invertible, and has an everywhere positive Jacobian.
Using this diffeomorphism, we can define a new field $\tilde{\Th}$ on $\tilde{M}$
by setting $\tilde{\Th} = \Th\circ \bm{\xi}$ where $\circ$ denotes composition.
 Our aim is to show that $\tilde{\Th}$
satisfies an equation of the same form as \eqref{eq:heat_differential} along with
its associated boundary condition. To do this, let $\tilde{U}$
be a sub-domain in $\tilde{M}$, and $U = \bm{\xi}(\tilde{U})$ its image
under the diffeomorphism. We recall that
the vector area elements on $\partial U$ and $\partial \tilde{U}$ are related by
\begin{equation}\label{eq:dStrans}
\bb{n}\dS = J_{\xi}\FxiinvT\cdot\tilde{\bb{n}}\,\mathrm{d}\tilde{S},
\end{equation}
where $\Fxi=\left(\nabla\bs{\xi}\right)\trans$ is the deformation gradient, and $J_{\xi}\coloneqq\det{\Fxi}$
the Jacobian of $\bm{\xi}$.  We can use $\bm{\xi}$ to transform the surface integral within \eqref{eq:heat_integral}  to become
\begin{align}\label{eq:heat_integral_trans1}
  \begin{split}
  \intdU{\inprod{\bb{k}\cdot\nabla \Th}{\bb{n}}} &= \intdUtilde{\Big\langle\left(
    \mathbf{k}\circ \bm{\xi}\right)\cdot\left(\FxiinvT\cdot\nabla\tilde{\Th}\right),J_{\xi}\FxiinvT\cdot\tilde{\bb{n}}\Big\rangle} \\
  &= \intdUtilde{\inprod{\tilde{\bb{k}}\cdot\nabla\tilde{\Th}}{\tilde{\bb{n}}}},
  \end{split}
\end{align}
where we have made use of the chain rule to obtain the relation
\begin{equation}\label{eq:gradT_trans}
  (\nabla\tilde{\Th})\ofx = \Fxi\ofx\cdot(\nabla\Th)\ofxix,
\end{equation}
and  have defined a new conductivity tensor on $\tilde{M}$ by
\begin{equation}\label{eq:conductivity_trans}
\tilde{\bb{k}}\ofx = J_{\xi}\ofx\Fxiinv\ofx\bb{k}\ofxix\FxiinvT\ofx, 
\end{equation}
which we note is symmetric and positive-definite as is physically required.
The transformed integral takes the same form as that in \eqref{eq:heat_integral}, and
so  the temperature field $\tilde{\Th}$ on $\tilde{M}$ satisfies 
\begin{equation}\label{eq:heat_integral_trans2}
  \intdUtilde{\inprod{\tilde{\bb{k}}\cdot\nabla\tilde{\Th}}{\tilde{\bb{n}}}} = 0,
\end{equation}
for all sub-domains $\tilde{U}\subseteq \tilde{M}$. We noted above that this
is equivalent to the temperature field satisfying the steady-state heat equation
\begin{equation}\label{eq:heat_differential_trans}
  \mathrm{div}\big(\tilde{\bb{k}}\cdot\nabla\tilde{\Th}\big)=0,
\end{equation}
along with the corresponding insulating boundary condition
\begin{equation}\label{eq:heat_boundary_trans}
    \inprod{\tilde{\bb{k}}\cdot\nabla\tilde{\Th}}{\tilde{\bb{n}}} = \bb{0},
\end{equation}
on $\partial \tilde{M}$. It follows that, provided the appropriate transformed conductivity \eqref{eq:conductivity_trans} is used, the  transformed temperature field $\tilde{\Th}$ satisfies a heat equation on $\tilde{M}$ of exactly the same form as the original temperature field.

To see the applicability of this form invariance property to invisibility, consider a diffeomorphism $\bm{\xi}$ from $M$ onto itself, such that $\xix = \bb{x}$ on $\dM$. If $\Th$ is a solution to the heat equation with conductivity $\bb{k}$, then $\tilde{\Th}$ is also a solution to the heat equation with conductivity $\tilde{\bb{k}}$
defined in \eqref{eq:conductivity_trans}.
If we consider an external observer only able to make observations of temperature on $\dM$, then since, by construction, $\Th=\tilde{\Th}$ on $\dM$, the observer is unable to distinguish between the two conductivities
$\bb{k}$ and $\tilde{\bb{k}}$. The space of such diffeomorphisms $\bm{\xi}$ is clearly infinite-dimensional, and so an inverse problem to determine $\bb{k}$ from surface observations of $\Th$ possesses an infinite-fold non-uniqueness.

\section{Thermoelasticity}\label{sec:thermoelasticity}

The theory of thermoelasticity provides a framework for modelling  coupled thermal and mechanical responses of solids. A brief overview of the required  principles and equations follows, largely based on the formulation of Gurtin \emph{et al.} \cite{Gurtin2010}.

\subsection{Kinematics}

To describe the kinematics of a solid body it is conventional to use a referential formulation
(the equivalent terms ``material'' or ``Lagrangian'' are used by
some authors). We let $M\subseteq \mathbb{R}^{n}$ denote a \emph{reference body} whose elements stand in a
one-to-one correspondence with the particles of the body, with the associated mapping
defining the chosen \emph{reference configuration}. 
It is assumed that $M$ is a compact subset of $\mathbb{R}^{n}$ with open interior and
smooth boundary $\partial M$. In what follows, a \emph{sub-body} will mean a subset of $M$
that also satisfies these conditions.

At a time $t$, the particle labelled by $\mathbf{x} \in M$ is located at the point
$\bm{\varphi}(\bb{x},t) \in \mathbb{R}^{n}$ of physical space, and this defines
a unique  mapping $\bm{\varphi}:M\times\mathbb{R}\rightarrow\mathbb{R}^n$
known as the \emph{motion} of the body relative to the reference configuration. The
body's \emph{instantaneous configuration} at time $t$ is the induced mapping
$\bm{\varphi}(\cdot,t):M\rightarrow\mathbb{R}^n$, which
is required to be a smooth orientation-preserving embedding. 
The referential velocity $\bb{v}\ofxt = \partial\bs{\varphi}/\partial t$ is
defined to be the velocity of the particle
labelled by $\mathbf{x}$  when, at time $t$, it is located at the spatial point $\bm{\varphi}\ofxt$.
The deformation gradient of the motion is defined by $\bb{F} = \left(\nabla\bs{\varphi}\right)\trans$.

\subsection{Balance laws}

The basic equations of thermoelasticity can be written as integral balance laws for linear momentum
and energy that must hold within an arbitrary sub-body, along with an inequality concerning the
production of entropy. The balance of linear momentum within a sub-body $U$ is given by
\begin{equation}\label{eq:linearmomentumbalance}
  \frac{\diff}{\diff t}\intU{\rho\bb{v}} = \intdU{\bb{T}\cdot\bb{n}} + \intU{\bb{b}},
\end{equation}
with $\rho$ the referential density, $\bb{T}$ the first Piola-Kirchhoff stress tensor, and $\bb{b}$ a given body force, defined per unit volume on $M$. 

To state the first law of thermodynamics, we introduce the internal energy per unit volume $\varepsilon$, the heat flux $\bb{q}$ and an external volumetric heat source $q$, all as measured on the reference body. The balance of energy within an arbitary sub-body $U$ can then be expressed as
\begin{equation}\label{eq:energybalance}
  \frac{\diff}{\diff t}\intU{\varepsilon} = \intU{\inprod{\bb{T}}{\dot{\bb{F}}}} - \intdU{\inprod{\bb{q}}{\bb{n}}} + \intU{q},
\end{equation}
where we denote by $\inprod{\mathbf{A}}{\mathbf{B}} = A_{ij}B_{ij}$ the usual inner product for
linear operators on $\mathbb{R}^{n}$, with the summation convention applied over repeated indices.

Finally, we require a statement of the second-law of thermodynamics
in the form of the Clausius-Duhem inequality. Introducing the absolute temperature $\Th>0$, along with the entropy density $\eta$ and the free energy density $\psi = \varepsilon-\eta\Th$, both  as measured per unit volume on $M$, the local form of the entropy inequality can be shown to be
\begin{equation}\label{eq:entropyinequality}
  \dot{\psi} + \eta\dot{\vartheta} - \inprod{\bb{T}}{\dot{\bb{F}}} + \frac{1}{\vartheta}\inprod{\bb{q}}{\nabla\vartheta} \leq 0.
\end{equation}
Here it is worth noting that a basic assumption of the theory is that
the absolute temperature is defined such that $\bb{q}/\vartheta$ represents the flux of entropy.
It follows, in particular, that energy and heat always flow in the same direction.

\subsection{Constitutive relations}
A set of constitutive relations is needed for a complete theory. Again following \cite{Gurtin2010},
we assume  constitutive functions exist such that
\begin{align}\label{eq:constitutivePsi}
\psi &= \Psi(\vect{F},\Th, \nabla \Th, \vx),\\
\label{eq:constitutiveT}
\vT &= \mathbcal{T}(\vF,\Th, \nabla \Th, \vx),\\
\label{eq:constitutiveEta}
\eta &= \mathcal{H}(\vF,\Th, \nabla \Th, \vx),\\
\label{eq:constitutiveQ}
\vq &= \mathbcal{q}(\vF,\Th, \nabla \Th, \vx).
\end{align}
Relations between these consitutive functions can be obtained using
the Coleman-Noll procedure \cite{Coleman1963}. This is done by substituting the
constitutive functions into the  entropy inequality \eqref{eq:entropyinequality}
and using the chain rule to obtain
\begin{align}\label{eqn:coleman}
  \begin{split}
    \bigg\langle &\pderiv{\Psi\args}{\vF} - \mathbcal{T}\args\,, \dot{\vF} \bigg\rangle \\&+ \bigg(\pderiv{\Psi\args}{\Th} - \mathcal{H}\args\bigg)\dot{\Th} \\ &+ \bigg\langle\pderiv{\Psi\args}{(\nabla\Th)}\,, \nabla\dot{\Th}\bigg\rangle + \frac{1}{\Th}\langle\mathbcal{q}\args\,,\nabla \Th\rangle \leq 0,
    \end{split}      
\end{align}
where 
\begin{equation}\label{eqn:matrixderiv}
\left(\pderiv{\Psi}{\vF}\right)_{ij} \coloneqq \pderiv{\Psi}{F_{ij}}.
\end{equation}
At a given point, the fields $\vF$, $\Th$, $\nabla \Th$ and their time derivatives can have arbitrarily prescribed values, so the coefficients for $\dot{\vF}$, $\dot{\Th}$ and $\nabla\dot{\Th}$  in \eqref{eqn:coleman} must vanish, giving the constraints
\begin{align}\label{eq:constraintPsi}
\psi &= \Psi(\vF,\Th,\vx),\\
\label{eq:constraintT}
\vT &= \mathbcal{T}(\vF,\Th,\vx) = \pderiv{\Psi(\vF,\Th,\vx)}{\vF},\\
\label{eq:constraintEta}
\eta &= \mathcal{H}(\vF,\Th\,\vx) = -\pderiv{\Psi(\vF,\Th,\vx)}{\Th},
\end{align}
where  the first  equality states that the free energy
is independent of the temperature gradient, 
along with the remaining heat-conduction inequality
\begin{equation}\label{eq:heatconductioninequality}
\langle\mathbcal{q}\args\,,\nabla\Th\rangle\leq 0.
\end{equation}
The theory is left with two independent constitutive functions, $\Psi$ and $\mathbcal{q}$.

The requirement of material frame-indifference adds additional constraints on these constitutive functions, as shown in \cite{Gurtin2010},
\begin{align}\label{eq:Psiindifference}
  \Psi(\vF,\Th,\vx) &= \Psi(\bb{Q}\vF,\Th,\vx) \\\label{eq:qindifference}
  \mathbcal{q}(\vF,\Th,\nabla\Th,\vx) &= \mathbcal{q}(\bb{Q}\vF,\Th,\nabla\Th,\vx),
\end{align}
where $\bb{Q}$ is an arbitrary rotation applied in physical space. The conditions in \eqref{eq:Psiindifference} and \eqref{eq:qindifference}, together with the relations \eqref{eq:constraintT} and \eqref{eq:constraintEta}, are sufficient to ensure that $\mathbcal{T}$ and $\mathcal{H}$ also satisfy material frame-indifference.

Using the integral balance laws for linear momentum and energy along with the  constitutive relations,
coupled systems of partial differential equations
governing the evolution of $\bm{\varphi}$ and $\Th$ can be obtained along with associated boundary conditions. For definiteness, these are
\begin{equation}\label{eq:linearmomentumlocal}
  \rho\partial_t\bb{v} - \Div\vT = \vb,
\end{equation}
\begin{equation}\label{eq:energybalancelocal}
  \partial_t(\psi+\eta\Th)  - \inprod{\vT}{\dot{\vF}} + \Div\vq = q,
\end{equation}
where we denote by $\Div$ the divergence of a linear operator. The corresponding traction-free and insulating boundary conditions on $\dM$ are, respectively,
\begin{equation}\label{eq:tractionfreelocal}
  \vT\cdot\vn = \bb{0},
\end{equation}
\begin{equation}\label{eq:insulatinglocal}
  \inprod{\vq}{\vn} = 0.
\end{equation}
For the aims of this paper, however, these local equations are not required, and it will be
sufficient to leave the equations of motion in the form of integral balance laws. Indeed,
as we have already seen in Sect. \ref{sec:heat},  this approach greatly simplifies study of
form invariance properties.

\subsection{Particle relabelling transformations}\label{sec:thermotrans}

The description  of thermoelasticity given above required the introduction of
a reference configuration, with the motion $\bm{\varphi}$, the absolute temperature
$\Th$, and the constitutive functions $\Psi$ and $\mathbcal{q}$ being defined in terms of the
associated reference body $M$.
Crucially, however, the choice of  reference configuration was left  implicit
within the theory, and can
be made in many different ways \cite{Antman1995}. It follows that
the equations of motion should be form invariant  with respect to
a change of  reference configuration, and we verify that this is the case below. 
Moreover, we  determine how the different fields and
material parameters of the body change under such a transformation.

Consider two reference configurations used to describe the same
physical body, with the associated motions denoted by
$\tilde{\bm{\varphi}}:\tilde{M}\times \mathbb{R}\rightarrow \mathbb{R}^{n}$,
and ${\bm{\varphi}}:{M}\times \mathbb{R}\rightarrow \mathbb{R}^{n}$, respectively.  A point $\mathbf{x} \in \tilde{M}$  acts as a label for
a unique particle in the body, with this particle being also labelled by
some $\mathbf{y}\in M$. This one-to-one correspondence between points in the
two reference bodies defines a mapping $\bm{\xi}:\tilde{M}\rightarrow M$ which we
assume to be an orientation-preserving diffeomorphism. In the terminology of Al-Attar \& Crawford \cite{Al-Attar2016}, the diffeomorphism $\bm{\xi}$ is a \emph{particle relabelling
transformation}, and by construction the two motions are related through
\begin{equation}
  \tilde{\bm{\varphi}}(\mathbf{x},t) = \bm{\varphi}[\bm{\xi}(\mathbf{x}),t], 
\end{equation}
for all $(\mathbf{x},t) \in \tilde{M}\times \mathbb{R}$. In terms of
this mapping, we   define  $\Fxi=\left(\nabla\bs{\xi}\right)\trans$ and
$J_\xi \coloneqq \det{\Fxi}$.
It will also be useful to introduce a shorthand notation for the composition of fields as follows.
For time-independent $f$, we set 
\begin{equation}\label{eq:tindependentcomposition}
\bcomp{f}\ofx\coloneqq f\ofxix,
\end{equation}
in the usual manner. If $f$ is instead time-dependent, then
\begin{equation}\label{eq:tdependentcomposition}
  \bcomp{f}\ofxt\coloneqq f\ofxixt.
\end{equation}
From this point onwards, we will suppress arguments where possible.

As already noted, both motions must map a fixed particle to the same point in physical space, so 
\begin{equation}\label{eq:phitrans}
\tilde{\vphi} = \comp{\vphi}.
\end{equation}
This immediately yields
\begin{equation}\label{eq:vtrans}
\tilde{\vv} = \comp{\vv},
\end{equation}
and
\begin{equation}\label{eq:Ftrans}
\tilde{\vF} = \bcomp{\vF}\,\Fxi.
\end{equation}
The two referential temperatures, $\tilde{\Th}$ and $\Th$, must give the same temperature
at a fixed particle, and so
\begin{equation}\label{eq:thetatrans}
\tilde{\Th} = \comp{\Th}.
\end{equation}
This is equivalent to the definition of the field
$\tilde{\Th}$ within Sect. \ref{sec:heat}, though here we see a
clear physical motivation for the condition.

Consider an arbitrary sub-body $\tilde{U}\subseteq\tilde{M}$, and its image $U\subseteq M$. The integral of $\rho$ over $U$ gives the mass of the corresponding sub-body in physical space, which must be independent of the choice of reference configuration. It follows that
\begin{equation}\label{eq:rhointegral}
\intU{\rho} = \intUtilde{\bcomp{\rho}J_\xi} = \intUtilde{\tilde{\rho}}.
\end{equation}
As $U$ was arbitrary, the transformation of the
referential density is seen to be
\begin{equation}\label{eq:rhotrans}
\tilde{\rho} = J_\xi\,\comp{\rho}.
\end{equation}
The same transformation law applies to all fields whose integral over a volume gives a physical quantity. By this we mean that the integral gives a property of the corresponding sub-body in physical space, which we require to take the same value, independent of the choice of reference configuration.

The transformation of the stress $\vT$ can instead be found by considering the total surface force (in physical space) on a sub-body, which should be independent of the choice of referential description. This is expressed, using \eqref{eq:dStrans}, as
\begin{equation}\label{eq:Tintegral}
\intdU{\vT\cdot\vect{n}}=  \intdUtilde{\bcomp{\vT}\cdot(J_\xi\FxiinvT\cdot\tilde{\vn})} =  \intdUtilde{\tilde{\vT}\cdot\tilde{\vn}},
\end{equation}
which implies that
\begin{equation}\label{eq:Ttrans}
\tilde{\vT} = J_\xi\bcomp{\vT}\FxiinvT.
\end{equation}
By the same argument as for density, the body force density transforms as 
\begin{equation}\label{eq:btrans}
\tilde{\vb} = J_\xi\,\comp{\vb}.
\end{equation}
We have established the transformations necessary to demonstrate the form invariance of the balance of linear momentum. Each integral in \eqref{eq:linearmomentumbalance} transforms to an integral of exactly the same form, giving the transformed equation 
\begin{equation}\label{eq:linmomtrans}
\frac{\diff}{\diff t}\intUtilde{\tilde{\rho}\,\tilde{\vv}} = \intdUtilde{\tilde{\vT}\cdot\tilde{\vn}} + \intUtilde{\tilde{\vb}}.
\end{equation}

Turning to the balance law for energy, the quantities $\psi$, $\varepsilon$, $\eta$ and $q$ can all be integrated over a volume to give physical quantities, so their transformations are given simply by
\begin{align}\label{eq:psitrans}
\tilde{\psi} &= J_\xi\comp{\psi},\\
\label{epsilontrans}
\tilde{\varepsilon} &= J_\xi\comp{\varepsilon},\\
\label{eq:etatrans}
\tilde{\eta} &= J_\xi\comp{\eta},\\
\label{eq:qscalartrans}
\tilde{q} &= J_\xi \comp{q}.
\end{align}
The integral of the heat flux over the boundary of a sub-body $U$ gives the heat flow out of $U$, which is a physical quantity. Using \eqref{eq:dStrans}, we can then find the transformation of $\vq$ by considering
\begin{equation}\label{eq:qintegral}
\intdU{\inprod{\vq}{\vn}}= \intdUtilde{\inprod{\comp{\vq}}{J_{\xi}\FxiinvT\cdot\tilde{\vn}}}= \intdUtilde{\inprod{\tilde{\vq}}{\tilde{\vn}}}, 
\end{equation}
and from this we obtain
\begin{equation}\label{eq:qtrans}
\tilde{\vq}=J_\xi\Fxiinv\cdot\bcomp{\vq}.
\end{equation}
Using the transformations \eqref{eq:Ftrans} and \eqref{eq:Ttrans}, we see that
\begin{equation}\label{eq:powertrans}
\inprod{\tilde{\vT}}{{\dot{\tilde{\vF}}}} = \tilde{T}_{ij}\dot{\tilde{F}}_{ij} = J_\xi T_{ik}[\FxiinvT]_{kj}\dot{F}_{il}[\Fxi]_{lj} = J_\xi T_{ik}\dot{F}_{il}\delta_{kl} = J_\xi\inprod{\vT}{\dot{\vF}}, 
\end{equation}
which leads to 
\begin{equation}\label{eq:powertransintegral}
\intU{\inprod{\vT}{\dot{\vF}}} = \intUtilde{\inprod{\comp{\vT}}{\comp{\dot{\vF}}}J_\xi} = \intUtilde{\inprod{\tilde{\vT}}{\dot{\tilde{\vF}}}}.
\end{equation}
Putting these results together, we see from  \eqref{eq:energybalance} that
\begin{equation}\label{eq:energytrans}
\frac{\diff}{\diff t}\intUtilde{\tilde{\varepsilon}} = \intUtilde{\inprod{\tilde{\vT}}{\dot{\tilde{\vF}}}} - \intdUtilde{\inprod{\tilde{\vq}}{\tilde{\vn}}} +\intUtilde{\tilde{q}},
\end{equation}
from which it follows  that balance of energy is indeed invariant under a
change of reference configuration.

To deal with the transformation of the entropy inequality, 
we need the identity
\begin{equation}\label{gradthetatrans}
\nabla\tilde{\Th} = \FxiT\cdot[\comp{(\nabla\Th)}].
\end{equation}
Combining this with \eqref{eq:qtrans}, we obtain
\begin{equation}\label{eq:entropyinequality1}
\frac{1}{\tilde{\Th}}\inprod{\tilde{\vq}}{\nabla\tilde{\Th}} = \frac{1}{\comp{\Th}}J_\xi\inprod{\Fxiinv\cdot\bcomp{\vq}}{\FxiT\cdot[\comp{(\nabla\Th)}]} = \frac{J_\xi}{\Th}\inprod{\vq}{\nabla\Th}.
\end{equation}
As each term in \eqref{eq:entropyinequality} simply picks up a factor $J_\xi$ on transformation, the integral form of the entropy inequality,
\begin{equation}\label{eq:entropyintegral}
\intU{\big(\dot{\psi} + \eta\dot{\vartheta} - \inprod{\vT}{\dot{\vF}} + \frac{1}{\vartheta}\inprod{\vq}{\nabla\Th}\big)} \leq 0,
\end{equation}
is transformed to
\begin{equation}\label{eq:entropyintegraltrans}
\intUtilde{\big(\dot{\tilde{\psi}} + \tilde{\eta}\dot{\tilde{\vartheta}} - \inprod{\tilde{\vT}}{\dot{\tilde{\vF}}} + \frac{1}{\tilde{\vartheta}}\inprod{\tilde{\vq}}{\nabla\tilde{\Th}}\big)} \leq 0,
\end{equation}
and so the second law of thermodynamics is still satisfied.

\subsection{Transformations of constitutive relations}
The constitutive functions on $\tilde{M}$ can now be expressed in terms of those on $M$, by observing that the quantities which they specify must transform in the appropriate manner. These transformations are, using \eqref{eq:psitrans} and \eqref{eq:qtrans}, 
\begin{equation}\label{eq:Psitrans}
\tilde{\Psi}(\tilde{\vF}, \tilde{\Th},\vx) = J_\xi \Psi\big[\comp{\vF},\comp{\Th},\bm{\xi}(\bb{x})\big] = J_\xi \Psi[\tilde{\vF}\Fxiinv, \tilde{\Th},\bm{\xi}(\bb{x})],
\end{equation}
\begin{align}\label{eqn:Qtrans}
  \begin{split}
    \tilde{\mathbcal{q}}(\tilde{\vF}, \tilde{\Th},\nabla\tilde{\Th},\bb{x}) &= J_\xi\Fxiinv\cdot \mathbcal{q}\big[\comp{\vF},\comp{\Th},\comp{(\nabla\Th)},\bm{\xi}(\bb{x})\big]\\ &= J_\xi\Fxiinv\cdot\mathbcal{q}[\tilde{\vF}\Fxiinv,\tilde{\Th},\FxiinvT\cdot\nabla\tilde{\vartheta},\bm{\xi}(\bb{x})].
    \end{split}
\end{align}
The chain rule can be used to show that \eqref{eq:Psitrans} is consistent with the transformations \eqref{eq:Ttrans} and \eqref{eq:etatrans} for $\vT$ and $\eta$.
Provided that $\Psi$ satisfies \eqref{eq:Psiindifference}, we see that
\begin{equation}\label{eq:Psitransinidifference}
  \tilde{\Psi}(\bb{Q}\tilde{\vF},\tilde{\Th},\vx) = J_{\xi}\Psi[\bb{Q}\tilde{\vF}\Fxiinv,\tilde{\Th},\bm{\xi}(\bb{x})] = J_{\xi}\Psi[\tilde{\vF}\Fxiinv,\tilde{\Th},\bm{\xi}(\bb{x})] = \tilde{\Psi}(\tilde{\vF},\tilde{\Th},\vx),
\end{equation}
meaning that the transformed constitutive relation also satisfies material frame-indifference. There is an identical result for $\tilde{\mathbcal{q}}$, if $\mathbcal{q}$ satisfies \eqref{eq:qindifference}. 

\subsection{Reduction to finite elasticity}\label{sec:reductionelasticity}
In the limit of no thermomechanical coupling and no thermal disturbance, these results reduce neatly to those of Al-Attar \& Crawford \cite{Al-Attar2016} and to the relevant parts of Yavari \& Golgoon \cite{Yavari2018},
whose work is presented in a  sophisticated geometrical language.
In particular, the transformations given by \eqref{eq:phitrans}, \eqref{eq:vtrans}, \eqref{eq:Ftrans}, \eqref{eq:rhotrans} and \eqref{eq:Ttrans} have direct counterparts. If the dependence on $\Th$ is removed, then the free energy transformation \eqref{eq:Psitrans} is seen to correspond to (71) in \cite{Al-Attar2016}. As in the case of finite elasticity, the non-uniqueness arising from form invariance does not imply invisibility, because the transformations are made between reference bodies, without changing the parameters of the physical body being described. As such, the mechanical deformation is seen to break the invisibility possessed by the pure heat diffusion problem.

\section{Linearised thermoelasticity}\label{sec:linthermo}
\subsection{Linearisation}\label{sec:linearisation}
In order to gain further insight into the implications of thermo-mechanical coupling for invisibility, we now look at the linearised theory of thermoelasticity. We consider small deviations from an equilibrium configuration
\begin{align}\label{eq:ulinear}
\vphi(\vx,t)&=\vphi_0(\vx) + \epsilon\vect{u}(\vx,t),\\
\label{eq:thetalinear}
\Th(\vx,t) &= \Th_0 + \epsilon(\Th(\vx,t)-\Th_0),
\end{align}
where $\Th_0$ is a constant temperature, $\vphi_0$ is a static equilibrium configuration relative to $M$ at this temperature, and $\epsilon$ is a bookkeeping parameter which can eventually be set to unity. In many applications, it is conventional to choose a reference configuration such that $\vphi_0(\vx) = \vx$. However, we need to treat the more general case, as we are considering transformations between reference configurations.

To obtain the linearised equations of motion, we simply expand the exact form
of the local balance laws to first order in $\epsilon$. The deformation gradient becomes 
\begin{equation}\label{eq:Flinear}
\vF(\vx,t) = \vF_0 (\vx) + \epsilon\vF_u (\vx,t),
\end{equation}
where $\vF_0$ and $\vF_u$ take their expected meanings as the deformation gradients of $\vphi_0$ and $\bb{u}$, respectively.
The stress can be expanded about the equilibrium point as
\begin{multline}\label{eqn:Tlinear}
  \vT = \mathbcal{T}[\vF_0 +\epsilon\vF_u, \Th_0 + \epsilon(\Th-\Th_0),\vx ] = \\\mathbcal{T}(\vF_0, \Th_0 ,\vx) + \pderiv{\mathbcal{T}}{\vF}[\vF_0, \Th_0, \vx]\cdot(\epsilon\vF_u) +
  \pderiv{\mathbcal{T}}{\Th}[\vF_0, \Th_0, \vx]\epsilon(\Th-\Th_0) + O(\epsilon^2) \\ = \vT_0 + \epsilon\big[\elast\cdot\vF_u + \vM(\Th-\Th_0)\big] + O(\epsilon^2),
\end{multline}
where we have defined the equilibrium stress
\begin{equation}\label{eq:T0}
\vT_0(\vx) = \mathbcal{T}( \vF_0, \Th_0, \vx),
\end{equation}
the elasticity tensor
\begin{equation}\label{eq:lambda}
[\elast(\vx)]_{ijkl} = \frac{\partial^2\Psi}{\partial F_{ij}\partial F_{kl}}[\vF_0, \Th_0, \vx],
\end{equation}
and the stress-temperature modulus
\begin{equation}\label{eq:M}
\vM(\vx) = \frac{\partial^2\Psi}{\partial\vF\partial\Th}[\vF_0, \Th_0, \vx].
\end{equation}
It can be shown \cite{Gurtin2010} that, due to the heat conduction inequality \eqref{eq:heatconductioninequality}, the most general form of the constitutive equation for heat flux is Fourier's law
\begin{equation}\label{eq:fourier}
\mathbcal{q}(\vect{F},\Th, \nabla \Th, \vx) = -\epsilon\vk(\vx)\cdot\nabla\Th + O(\epsilon^2),
\end{equation}
with $\vk$ a positive-definite conductivity tensor.

Substituting these expressions into the local linear momentum balance gives, to zeroth order,
\begin{equation}\label{eq:equilibriumcondition}
\mathrm{Div}\vT_0 = 0,
\end{equation}
which is simply the condition for equilibrium.
Collecting the first-order terms and neglecting any body forces, we arrive at the linearised
equation of motion
\begin{equation}\label{eq:lineom}
\rho\partial_t^2\vu = \Div[\elast\cdot\vF_u+\vM(\Th-\Th_0)].
\end{equation}
It can similarly be shown that the local linearised energy balance, neglecting any external heat sources, takes the form
\begin{equation}\label{eq:linenergy}
c\dot{\Th} = \mathrm{div}(\vk\cdot\nabla\Th) + \Th_0\inprod{\vM}{\dot{\vF}_u},
\end{equation}
where
\begin{equation}\label{eq:heatcapacity}
c = -\Th_0\frac{\partial^2\Psi}{\partial\Th^2}[\vF_0, \Th_0, \vx],
\end{equation}
is the heat capacity measured per unit volume in the reference body.

\subsection{Particle relabelling transformations in linearised thermoelasticity}

Under a particle relabelling transformation, the linearised equations should be form invariant, as they are derived from the form invariant equations of finite thermoelasticity. The transformations of the linearised moduli are readily derived from the transformations of the constitutive functions, \eqref{eq:Psitrans} and \eqref{eqn:Qtrans}. They are
\begin{align}\label{eq:lambdatrans}
\tilde{\Lambda}_{ijkl} &= J_\xi[\Fxiinv]_{jm}[\Fxiinv]_{ln}\comp{\Lambda_{imkn}},\\
\label{eq:Mtrans}
\tilde{\vM} &= J_\xi\bcomp{\vM}\FxiinvT,\\
\label{eq:ctrans}
\tilde{c} &= J_\xi\,\comp{c},\\
\label{eq:ktrans}
\tilde{\vect{k}} &=\Fxiinv\bcomp{\vect{k}}\FxiinvT.
\end{align}
These transformations are indeed consistent with the form invariance of \eqref{eq:lineom} and \eqref{eq:linenergy}.

\subsection{Weak thermomechanical coupling}

The linearised equations of thermoelasticity, \eqref{eq:lineom} and \eqref{eq:linenergy}, broadly resemble, respectively, the elastodynamic equation of motion and the heat equation. There is only a single term in each which couples mechanics with thermodynamics. In the limit $\vM\rightarrow\vect{0}$, the two decouple completely. We now consider the case in which the coupling between mechanics and thermodynamics is weak, in order to gain insight into the nature of invisibility in thermoelasticity.

If the small degree of coupling is here indicated by a small parameter $\epsilon$,  then the linearised equations are
\begin{equation}\label{eq:lineomsmall}
\rho\partial_t^2\vu = \Div[\elast\cdot\Fu+\epsilon\vM(\Th-\Th_0)],
\end{equation}
\begin{equation}\label{eq:linenergysmall}
c\dot{\Th} = \mathrm{div}(\vk\cdot\nabla\Th) + \epsilon\Th_0\inprod{\vM}{\dot{\vF}_{u}}.
\end{equation}
Note that this is a different meaning to the $\epsilon$ of Sect. \ref{sec:linearisation}.

Consider an initial value problem driven by a temperature disturbance on $M$, and on the time interval $[0,T]$. Let the initial conditions be
\begin{equation}\label{eq:initialtemp}
  \Th(\vx,0) = \hat{\Th}(\vx),\qquad \left|\frac{\hat{\Th}(\vx)-\Th_0}{\Th_0}\right|\ll 1,
\end{equation}
\begin{equation}\label{eq:initialdisplacement}
  \vu(\vx,0)=\partial_t\vu(\vx,0) = 0,
\end{equation}
with $\hat{\Th}$ some spatially varying initial temperature distribution. We additionally impose the traction-free and insulating boundary conditions
\begin{equation}\label{eq:boundaryconditions}
  [\elast\cdot\vF_u + \epsilon\vM(\Th-\Th_0)]\cdot\vn = \bb{0}, \qquad \inprod{\mathbf{k}\cdot
    \nabla\Th}{\vn} = 0, \quad \vx \in \partial M.
\end{equation}
To look for approximate solutions to this problem, we
perturbatively expand the dynamic fields in powers of $\epsilon$,
\begin{align}\label{eq:thetaexpansion}
  \Th-\Th_0 = (\Th\zero-\Th_0) + \epsilon\Th\one + \epsilon^2\Th\two + \dots,\\
\label{eq:uexpansion}
  \vu = \vu\zero + \epsilon\vu\one + \epsilon^2\vu\two + \dots,
\end{align}
and define $\Fu^{(j)} \coloneqq (\nabla\vu^{(j)})\trans$. The zeroth-order equations are seen to be
\begin{align}\label{eq:eom0}
\rho\,\partial_t^2\vu\zero = \Div(\elast\cdot\Fu\zero),\\
\label{eq:energy0}
  c\,\dot{\Th}\zero = \div(\vk\cdot\nabla\Th\zero),
\end{align}
with initial and boundary conditions
\begin{align}\label{eq:initialconditions0}
  \Th\zero(\vx,0) = \hat{\Th}\ofx, \quad \vu\zero(\vx,0) = \partial_t\vu\zero(\vx,0) = 0,\\
\label{eq:boundaryconditions0}
  (\elast\cdot\Fu\zero)\cdot\vn = 0,\quad\inprod{\mathbf{k}\cdot\nabla\Th\zero}{\vn} =0,\quad\vx\in\dM.
\end{align}

These equations clearly imply that $\vu\zero = \bb{0}$, while $\Th\zero$ is the unique solution to the heat diffusion equation, depending on the parameters $c$, $\vk$ and $\hat{\Th}$. In the case $\epsilon = 0$, this is the complete solution to the problem as the thermal and mechanical responses are completely decoupled. The corresponding inverse problem, as considered in Sect. \ref{sec:heat}, would be to determine $(c,\vk,\hat{\Th})$ from observations of $\Th$ on $\dM$ over the interval $[0,T]$. Consider a diffeomorphism $\bm{\xi}:M\rightarrow M$, with $\bs{\xi}\ofx = \vx$ for $\vx\in\dM$. $\tilde{\Th}\zero = \comp{\Th\zero}$ is a solution to the transformed equation
\begin{equation}\label{eq:heattransformed2}
\tilde{c}\dot{\tilde{\Th}}\zero = \mathrm{div}(\tilde{\vk}\cdot\nabla\tilde{\Th}\zero),
\end{equation}
subject to the conditions
\begin{equation}\label{eq:initialtemptrans}
\tilde{\Th}\zero(\vx,0) = \tilde{\hat{\Th}}\ofx = \bcomp{\hat{\Th}}\ofx, \qquad \nabla\tilde{\Th}\zero\cdot\vn=0, \quad \vx\in\partial M.
\end{equation}
As in Sect. \ref{sec:heat}, it is impossible to distinguish between the parameters $(c,\vk,\hat{\Th})$ and $(\tilde{c},\tilde{\vk},\tilde{\hat{\Th}})$ because $\tilde{\Th}\zero = \Th\zero$ on $\dM$. This implies thermal invisibility.

Returning to the case of $\epsilon\neq 0$, the first-order equations are
\begin{align}\label{eq:eom1}
  \rho\,\partial_t^2\vu\one &= \Div[\elast\cdot\Fu\one + \bb{M}(\Th\zero-\Th_0)],\\
\label{eq:energy1}
  c\,\dot{\Th}\one &= \div(\vk\cdot\Th\one),
\end{align}
with initial and boundary conditions
\begin{align}\label{eq:initialconditions1}
  &\Th\one(\vx,0) = 0, \quad \vu\one(\vx,0) = \partial_t\vu\one(\vx,0) = 0,\\
\label{eq:boundaryconditions1}
&(\elast\cdot\Fu\one)\cdot\vn = -[\bb{M}(\Th\zero-\Th_0)]\cdot\vn,
\quad\inprod{\mathbf{k}\cdot\nabla\Th\one}{\vn} =0,\quad\vx\in\dM.
\end{align}
This implies $\Th\one = 0$, while $\vu\one$ has some non-zero solution.
Further, the second-order equations can be derived and solved to find $\Th\two$ and $\vu\two$,
which are, in general, non-zero.

Given the absence of a first order perturbation to the temperature from thermomechanical coupling, it is illuminating to consider a transformation of the type discussed above, made on only the thermal parameters:
\begin{equation}\label{eq:thermalparameterstrans}
(\rho, \elast, \vM, c, \vk, \hat{\Th}) \mapsto (\rho, \elast, \vM, \tilde{c}, \tilde{\vk}, \tilde{\hat{\Th}}).
\end{equation}
The solution to the transformed problem takes the form
\begin{equation}\label{eq:solutiontrans}
  \tilde{\Th} = \tilde{\Th}\zero + \epsilon^2\tilde{\Th}\two = \comp{\Th\zero} + \epsilon^2\tilde{\Th}\two,
\end{equation}
with $\tilde{\Th}\two \ne \comp{\Th\two}$. On $\dM$, the zeroth-order part of the temperature field is unchanged by the transformation, so
\begin{equation}\label{eq:epsilonsquared}
  \tilde{\Th} - \Th = \epsilon^2(\tilde{\Th}\two - \Th\two) \sim O(\epsilon^2).
\end{equation}
Measurements made on the boundary are indeed altered by the transformation, but only by a term
of second-order in the coupling parameter. This means that for such thermally driven problems,
the invisibility of the pure heat diffusion problem is broken by coupling to mechanics, but only to second order in the degree of the coupling. It follows that so long as  the coupling is sufficiently small,
the heat diffusion can be said to retain \emph{approximate invisibility}.

\section{Conclusions}\label{sec:conclusions}
We have demonstrated the form invariance of the equations of thermoelasticity
under a change of reference configuration in the cases of both finite and linearised
disturbances, and used their linearised form to draw conclusions about the effect of
mechanical deformation on thermal invisibility. In the general case, the theory of
thermoelasticity does not allow for invisibility. However, we have arrived at the notion
of approximate invisibility, which is exhibited in materials with weak thermomechanical
coupling.

An interesting possibility is that a similar effect might occur if
elasticity is coupled to electromagnetism. Indeed, it is now well known that
the equations of electromagnetism allow for the possibility of perfect invisbility
cloaks, but this theory neglects any mechanical coupling. The extension of our results
to account for electromagnetic coupling is not, however, straightforward because
satisfactory continuum theory of electromagnetic materials has not been clearly
established \cite{Steigmann2009,Ericksen2007}. Nonetheless, it seems reasonable to conjecture that similar transformations can be made, as suggested in \cite{Hutter1975}. We suspect that coupling of the electromagnetic equations to those of elasticity will for similar reasons preclude the existence of perfect cloaks, although an approximate cloak may be possible if the coupling between electromagnetism and mechanics is weak.


\section*{Acknowledgements}
FS is supported by a Natural Environment Research Council studentship and a CASE award from Schlumberger Cambridge Research.


%
%
%
%
%
%

\bibliographystyle{ieeetr}       
\bibliography{thermomod}   


%
%

\end{document}